\documentclass[runningheads]{cl2emult}

\usepackage{makeidx}  % allows index generation
\usepackage{graphicx} % standard LaTeX graphics tool
\usepackage{subeqnar} % subnumbers individual equations
\usepackage{multicol} % used for the two-column index
\usepackage{cropmark} % cropmarks for pages without
\usepackage{math}     % placeholder for figures
\makeindex            % used for the subject index
\begin{document}

\title*{Are $EUR$ and $GBP$ different words for the same currency ?}
 \titlerunning{Are $EUR$ and $GBP$ different words}

\author{Marcel Ausloos\inst{1} \and Kristinka Ivanova\inst{2}}

\authorrunning{M. Ausloos and K. Ivanova }

\institute{GRASP, B5, University of Li$\grave e$ge, B-4000 Li$\grave e$ge, 
Belgium \and
Pennsylvania State University, University Park PA 16802, USA}

\maketitle

\begin{abstract} 
The British Pound ($GBP$) is not part of the Euro ($EUR$) monetary system. In
order to find out arguments on whether $GBP$ should join the $EUR$ or not
correlations are calculated between $GBP$ and $EUR$, including a reconstructed
$EUR$ for the time interval from 1993 till June 30, 2000. The distribution of
fluctuations of the exchange rates is Gaussian for the central part of the
distribution, but has fat tails for the large size fluctuations. 
Within the {\it
Detrended Fluctuation Analysis} ($DFA$) statistical method the power law
behaviour describing the root-mean-square deviation of the exchange rate
fluctuations is obtained as a function of time for the time interval 
of interest.
The time-dependent exponent evolution of the exchange rate 
fluctuations is given.
Statistical considerations imply that the $GBP$ is already behaving as a true
$EUR$. 

\end{abstract}

\noindent
{\bf Key words.} Econophysics; Detrended fluctuation analysis; Foreign
currency exchange rate; Euro; Scaling hypothesis

\section{Historical Introduction}

Inspired marketing may be needed to get the British to accept the euro
\cite{tomkins}. The Prime Minister in early February 2001 said that 
he would call
for a referendum on joining the euro within two years of a Labour 
party election
victory. The British public seems hostile. Between a Mori poll a 
couple of years
ago, and one in November 2000, the no's climbed from 56\% to 71\%. The big
problem for the euro camp is the emotional argument (abolishing the 
sterling). It
is very difficult to sell $advantages$ through rational arguments. It 
seems that
the people are shouting for a football team, and the business elite wish to
surrender sovereignty, since anyway they go for Tuscany for holidays, says Mr.
Byrne, chief executive of Weber Shandwick Public Affairs, as quoted by R.
Tomkins.\cite{tomkins} To argue in favor of the euro might be a hard 
task, but it
is simply true that the Britons will use the euro, whatever their 
taste, in 2005,
whether they like it or not. Mr. Draper, as quoted by R. Tomkins in
Ref. 1 considers that it has to be emphasized that 
it makes no
difference if all pounds turned into euro.\cite {tomkins} It is just 
another name
for money. This paper shows just $that$ from a currency exchange point of view.
So what the fuss?

Of course, from a financial and monetary policy, one can wonder what 
governments
should do? In 1961 Mundell\cite{Mundell1961} asked whether it is 
advantageous to
relinquish monetary sovereignty in favor of a common currency? For 
fixed exchange
rates, the central banks must intervene on the currency market in order to
satisfy the public demand for foreign currency at this exchange rate. As a
result, the central banks loose some control of the money supply which adjusts
itself to the domestic liquidity. To implement independent national monetary
policy by means of so-called {\it open market operations} becomes 
futile: neither
the interest rate nor the exchange rate can be affected. A currency union as
early as 1960 with fixed exchange rates existed within the so-called Bretton
Woods System. International capital movements were highly curtailed, in
particular by extensive capital and exchange rate controls. It was noticed that
due to high capital mobility in the world economy, such regimes with a
temporarily fixed, but adjustable, exchange rate were not robust. 
Whence several
governments have some fear about joining a system like $EUR$. Often 
psychological
or demagogic arguments are more relevant to the public than real 
economic ones, -
see the case of Denmark. The Blair Government supports the principle of joining
the single currency, if that is in the national economic interest but considers
{\it not to be ready yet}. Can the exchange rates serve as arguments? We have
searched for statistical correlations between $EUR$ and $GBP$ currency
fluctuations.

Hereby the question is raised on how the $GBP$ has fluctuated with respect to a
$false$ $EUR$, made of an equally weighted linear superposition of the 11
currencies forming $EUR$, and how $GBP/EUR$ fluctuates since Jan. 01, 1999.
Essentially it has been first searched whether the fluctuations did 
change after
the formal introduction of $EUR$. Next it has been observed whether there is an
{\it a posteriori} mathematical evidence for fluctuation correlations between
$GBP$, $EUR$ and the still presently national currencies. The considered time
interval is Jan.01, 1993 to June 30, 2000 is presented.

The  {\it Detrended Fluctuation Analysis} (DFA) is used. The $\alpha$ exponent
characterizing the power law over the longest possible scaling range pertaining
to this study is obtained. Next the local DFA study is performed and the
resulting so-called correlation matrix is presented, eliminating the time as a
series parameter. The data is summarized through the mean, the variance and the
median of the local $\alpha$ exponents. Some {\it financial policy 
statement and
historical considerations} arising from our observations serve as conclusions.

\section{Experimental Data}

The conversion rates of the $EUR$ participating countries were fixed 
by political
agreement based on the bilateral market rates of December 31, 1998. Using these
rates,\cite{quoteEUR} one Euro ($EUR$) can be represented as a weighted sum of
the eleven currencies $C_i$, $i= 1, 10$:

\begin{equation} 1 EUR = \sum_{i=1}^{10} (\delta_{i,2} + 1) \frac{\gamma_i}{11}
\, C_i \end{equation}

\noindent where $\gamma_i$ are the conversion rates and $C_i$ denote the
respective currencies, i.e. Austrian Schilling ($ATS$, $i$=1), Belgian Franc
($BEF$, $i$=2), German Mark ($DEM$, $i$=3), Spanish Peseta ($ESP$, $i$=4),
Finnish Markka ($FIM$, $i$=5), French Franc ($FRF$, $i$=6), Irish Pound ($IEP$,
$i$=7), Italian Lira ($ITL$, $i$=8), Dutch Guilder ($NLG$, $i$=9), Portuguese
Escudo ($PTE$, $i$=10). In view of the financial identity of the 
Luxemburg Franc
($LUF$), with the Belgian Franc ($BEF$), the latter is weighted by a factor of
two, whence the $\delta$ Kronecker symbol in the above equation. In order to
study correlations in the $EUR$/$GBP$ exchange rate, the $EUR$ existence can be
artificially extended backward, i.e., before Jan. 01, 1999 and thereby defining
an artificial $EUR$ before its birth.\cite{maki3} A data series of 
$EUR$ exchange
rates with respect to $GBP$ is constructed following the linear superposition
rule:

\begin{equation} 1 EUR/GBP = \sum_{i=1}^{10} (\delta_{i,2} + 1)
\frac{\gamma_i}{11} \, (C_i/GBP) \end{equation}

Since the number of data points of the exchange rates for the period starting
Jan. 1, 1993 and ending Dec. 31, 1998 is different for the eleven 
currencies, due
to different national and bank holidays a linear interpolation has 
been used for
the days when the banks are closed and official exchange rates are 
not defined in
some countries. The number $N$ of data points as equalized is $N = 1902$,
spanning the time interval from January 1, 1993 till June 30, 2000.
\footnote{This last day was chosen for the studies in order to remain coherent
and avoid a possible spurious effect arising from the Greek Drachma ($GRD$),
introduced as a supplementary currency in $EUR$ on June 19, 2000.}

The normalized and true $EUR$/$GBP$ exchange rates so reconstructed 
are given in
Fig. 1(a-b). For normalization purpose of the exchange rate Oct. 2, 
1996 has been
chosen as a typical day, i.e. are given in Table 1.

\begin{figure} \centering \includegraphics[width=.48\textwidth]{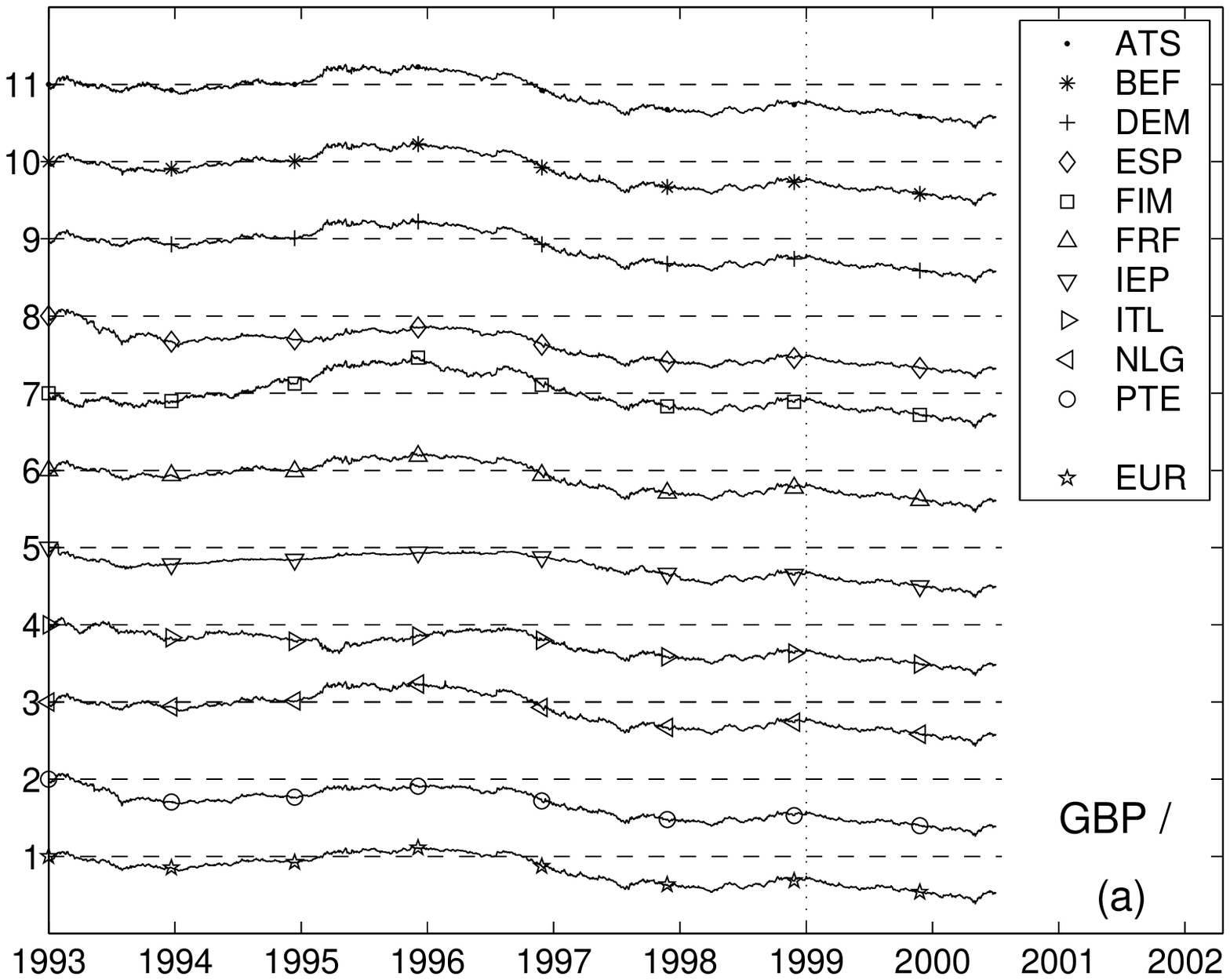}
\vfill \includegraphics[width=.48\textwidth]{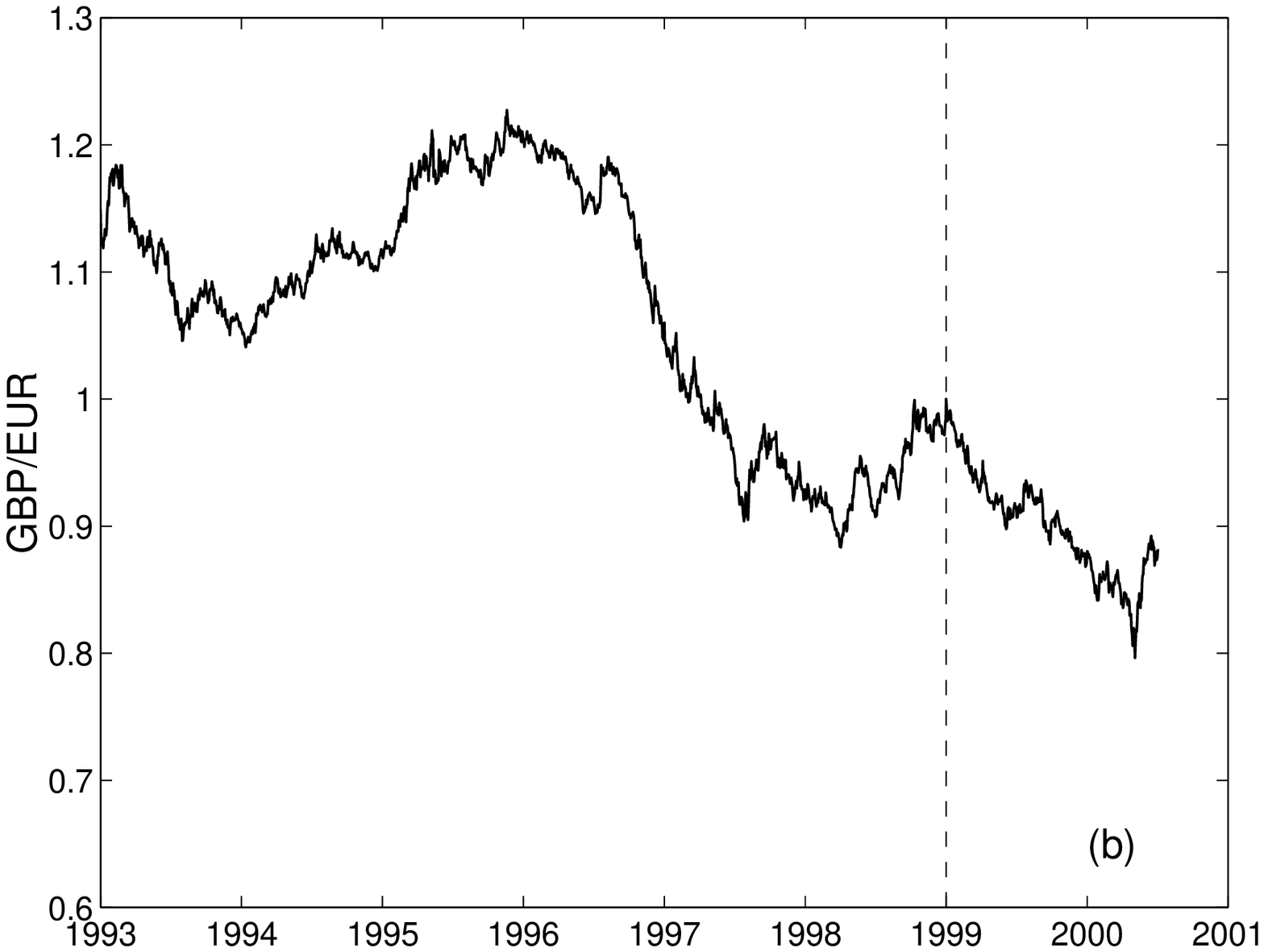} 
\caption{{\bf (a)} Normalized and {\bf (b)} true $EUR$ and
currencies forming the $EUR$ exchange rates with respect to $GBP$ between Jan.
01, 1993 and June 30, 2000.}
\label{eps2} \end{figure}

\begin{table}[ht] \caption{Indicative values, for normalization 
purposes of $GBP$
exchange rate on Oct 2, 1996,  e.g. 1 $GBP$ $\simeq$ 16.87 $ATS$. Numerical
values of DFA-$\alpha$ exponent for $EUR$-forming currency exchange rates. The
scaling time interval is $ca.$ one year. Values of the exponent of 
the power-law
of tail of the distribution of fluctuations for $EUR$-forming currency exchange
rates.} \begin{center} \begin{tabular}{||c||r|r|r|r|r|r|r|r|r|r||r||} \hline
\hline $C_i $ & ATS & BEF & DEM & ESP & FIM & FRF & IEP & ITL & NLG & PTE & EUR
\\ \hline ExR & 16.87 & 49.383 & 2.3976 & 201.59 & 7.1587 & 8.1241 & 0.9792 &
2372.0 & 2.6903 & 243.61 & 1.2234 \\ \hline $\alpha$ &$0.47 $& $0.47 
$& $0.50 $&
$0.51 $& $0.46 $& $0.47 $& $0.41 $& $0.46 $& $0.49 $& $0.45 $& $0.46 $ \\ &$
\pm0.02$& $ \pm0.02$& $ \pm0.02$& $ \pm0.02$& $ \pm0.02$& $ \pm0.02$& 
$ \pm0.03$&
$ \pm0.02$& $ \pm0.02$& $ \pm0.02$& $ \pm0.02$ \\ \hline $\mu$ 
&$4.10$ & $3.30$ &
$3.85$ & $3.67$ & $ 3.98$ & $3.98$ & $4.65$ & $3.30$ & $3.98$ & $3.56$ & $4.02$
\\ \hline \end{tabular} \end{center} \end{table}

It is known that ExR like $EUR/CHF$ and $EUR/DKK$ are pretty stable across the
transition to the $EUR$. However it seems that the ExR with respect 
to $GBP$ has
been much sensitive to the transition, with a noticeable decay of the 
$EUR$ value
after Jan. 01, 1999. See the 1995 bump of $ITL$, and dips at the end 
of 1993 for
$ATS$, $BEF$, $DEM$ and $NLG$ ExR with respect to $GBP$, the rate 
evolution being
rough.  The $ESP$, $FIM$, and $ITL$ seem currencies following weakly the ExR
majority (and the ExR statistical mean) evolution.

\section{Distribution of the fluctuations}

The distributions of the exchange rate fluctuations for $EUR/GBP$ are shown in
Fig. 2(a) for the time interval of interest. Each distribution 
central part, i.e.
the smallest fluctuations, is close to a Gaussian. The tails of the
distributions, i.e. the large fluctuations, strictly depart from the normal
distribution. Such tails usually have a slope markedly different from $-2$.
\cite{gopi} These so-called fat tails are found to follow a power-law
distribution with a slope of order of 4.0 for $EUR/GBP$. The distribution of
fluctuations in exchange rates for each 11 currency of interest is 
shown in Fig.
2(b-k). The characteristic power-law exponent of the tail of the distributions
for each $C_i/GBP$ is given in Table 1. The distribution for $DEM$, 
$FRF$, $FIM$,
$NLG$, and $ATS$  are close to that of the $EUR$, but that for $BEF$, $IEP$ and
$ITL$ is wide.  The $FRF$ large rate fluctuations are similar to those of the
$GBP$.

\begin{figure} \centering \includegraphics[width=.30\textwidth]{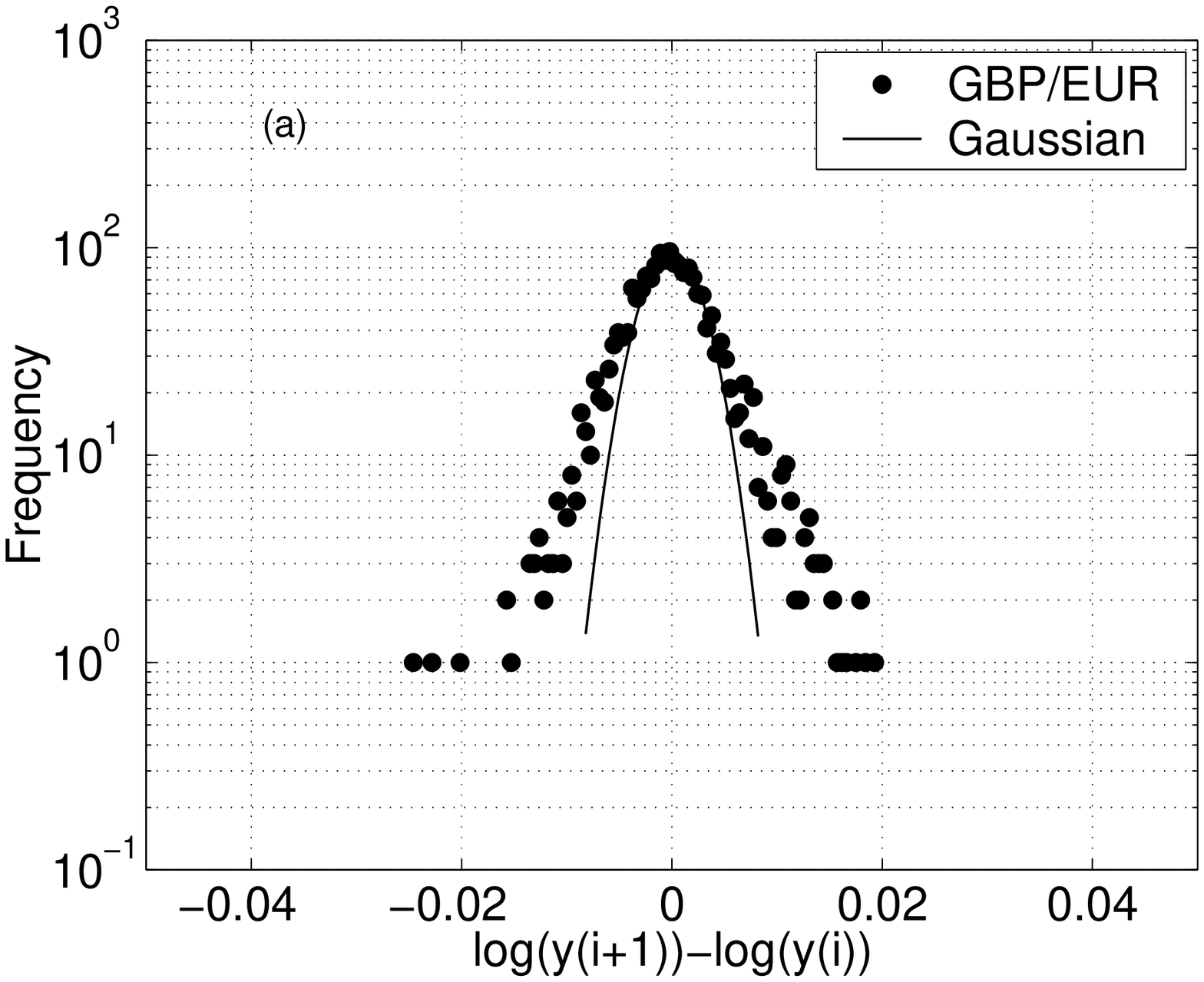}
\hfill \includegraphics[width=.30\textwidth]{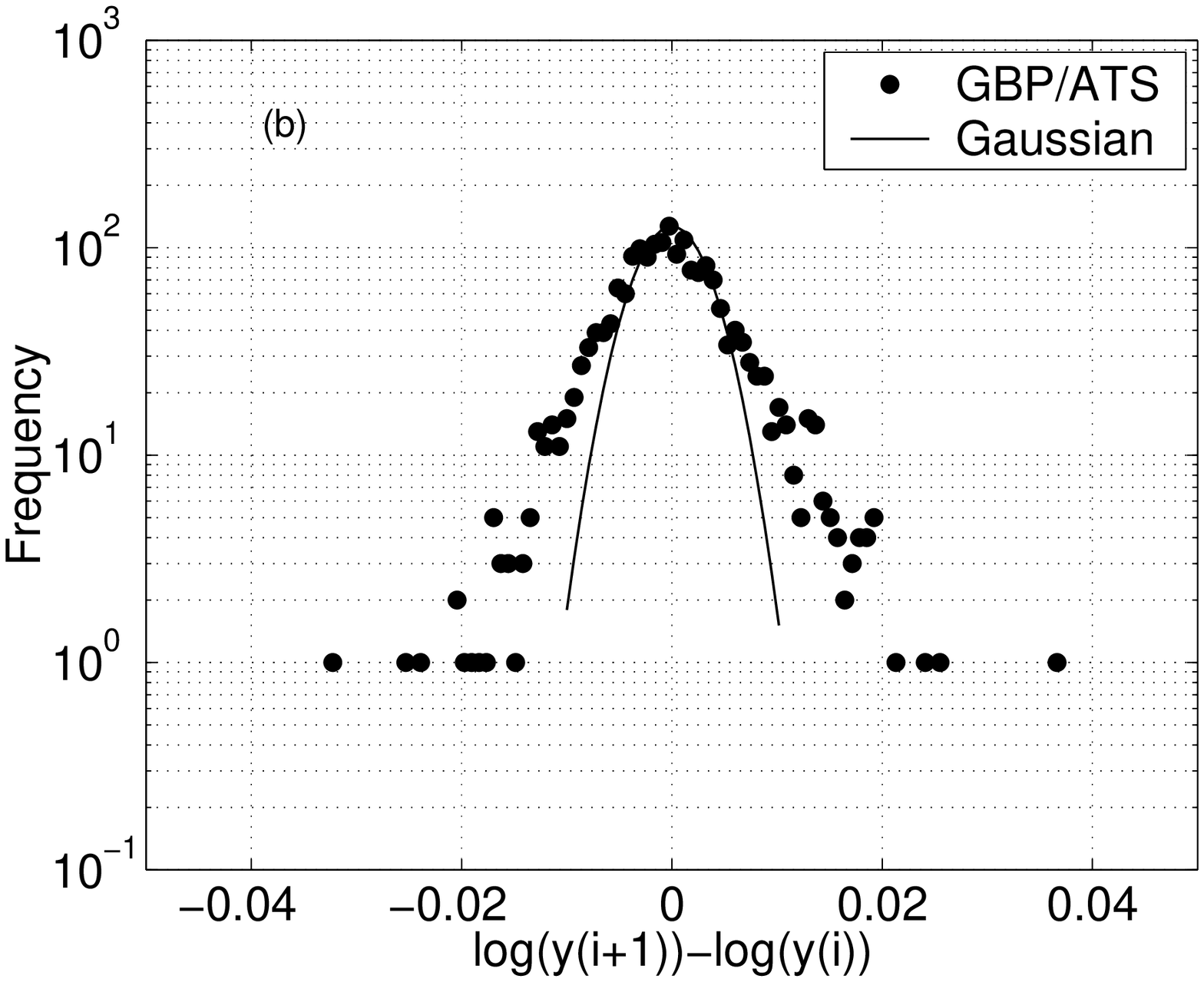} 
\vfill \includegraphics[width=.30\textwidth]{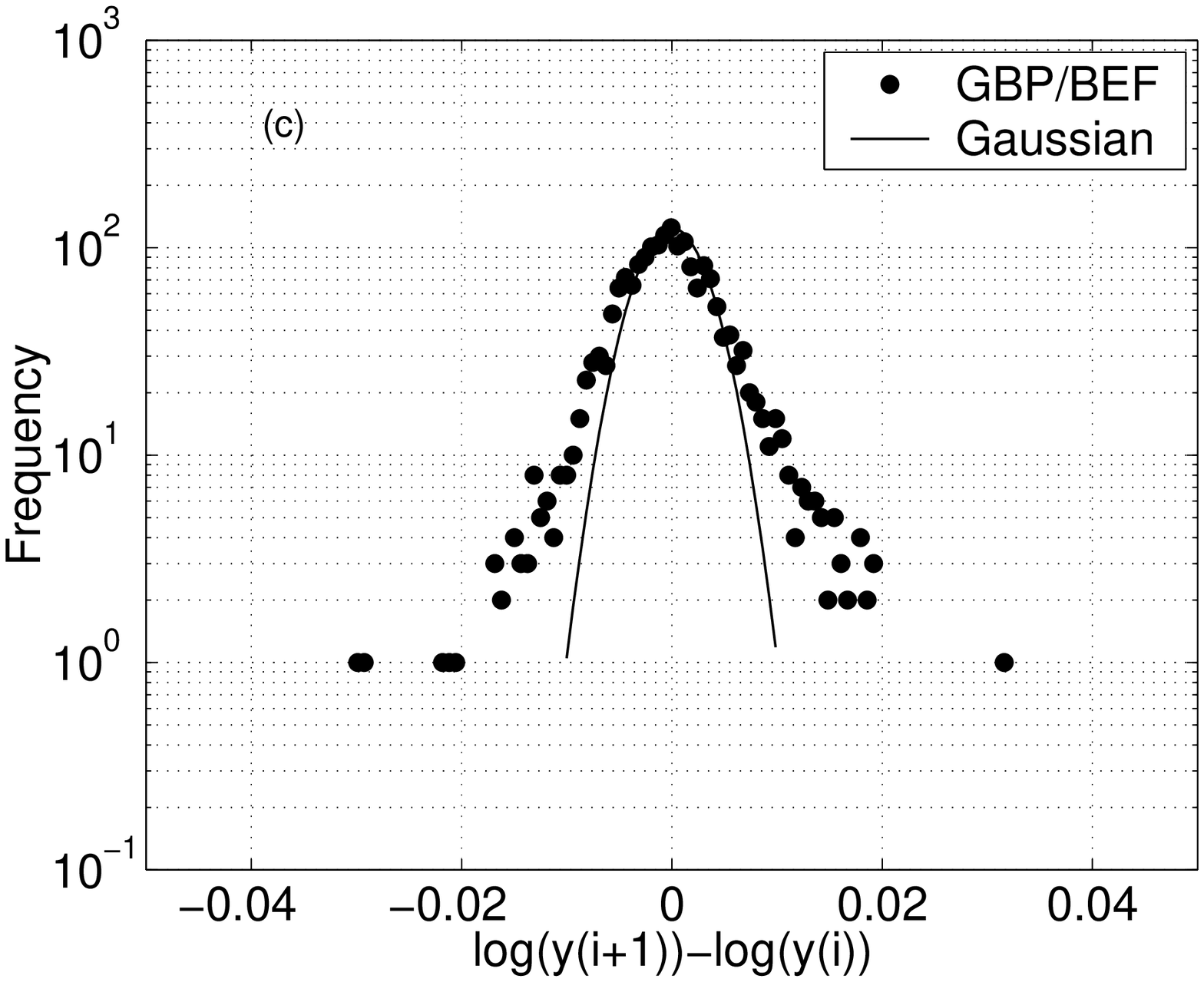} 
\hfill \includegraphics[width=.30\textwidth]{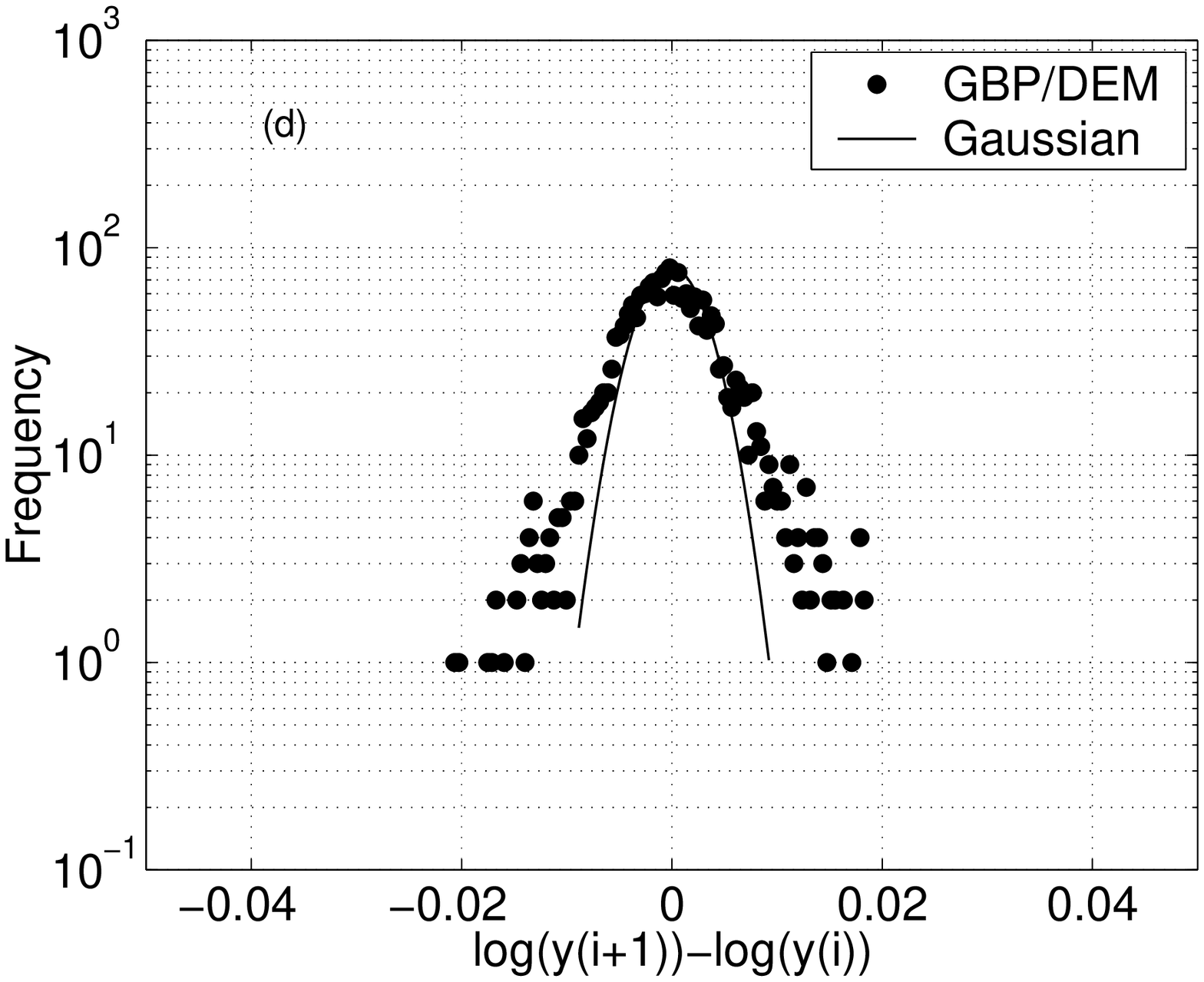} 
\hfill \includegraphics[width=.30\textwidth]{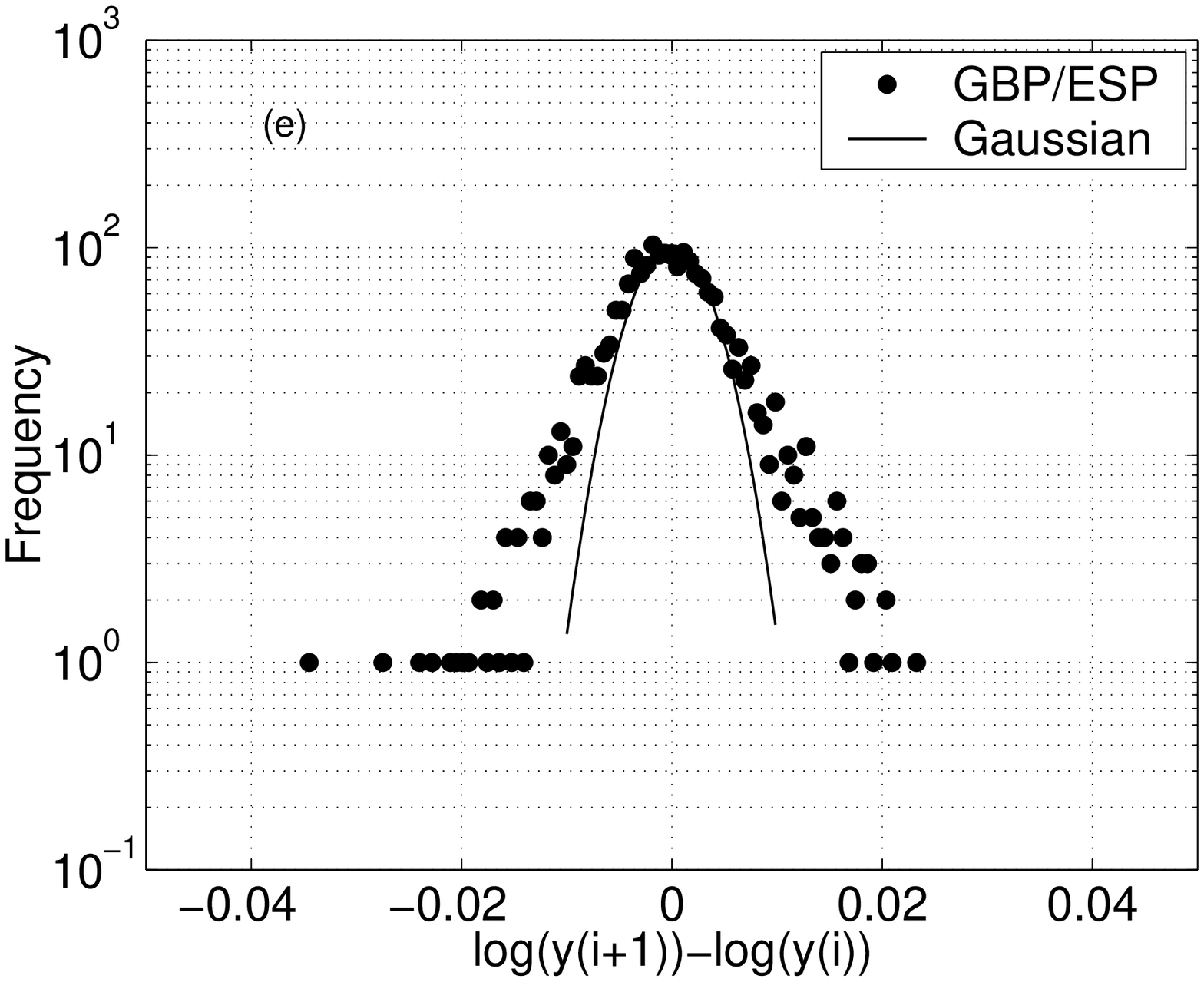} 
\vfill \includegraphics[width=.30\textwidth]{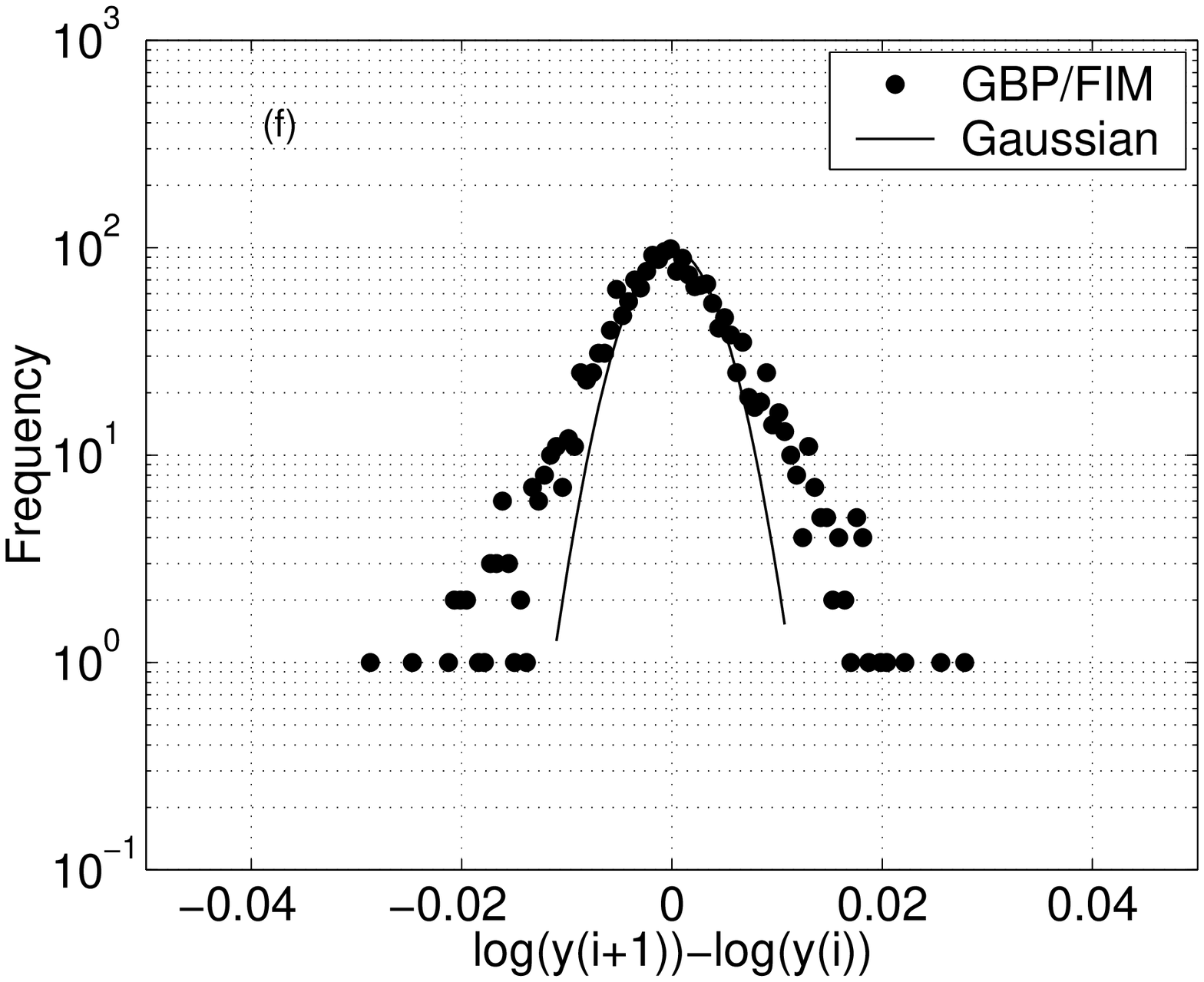} 
\hfill \includegraphics[width=.30\textwidth]{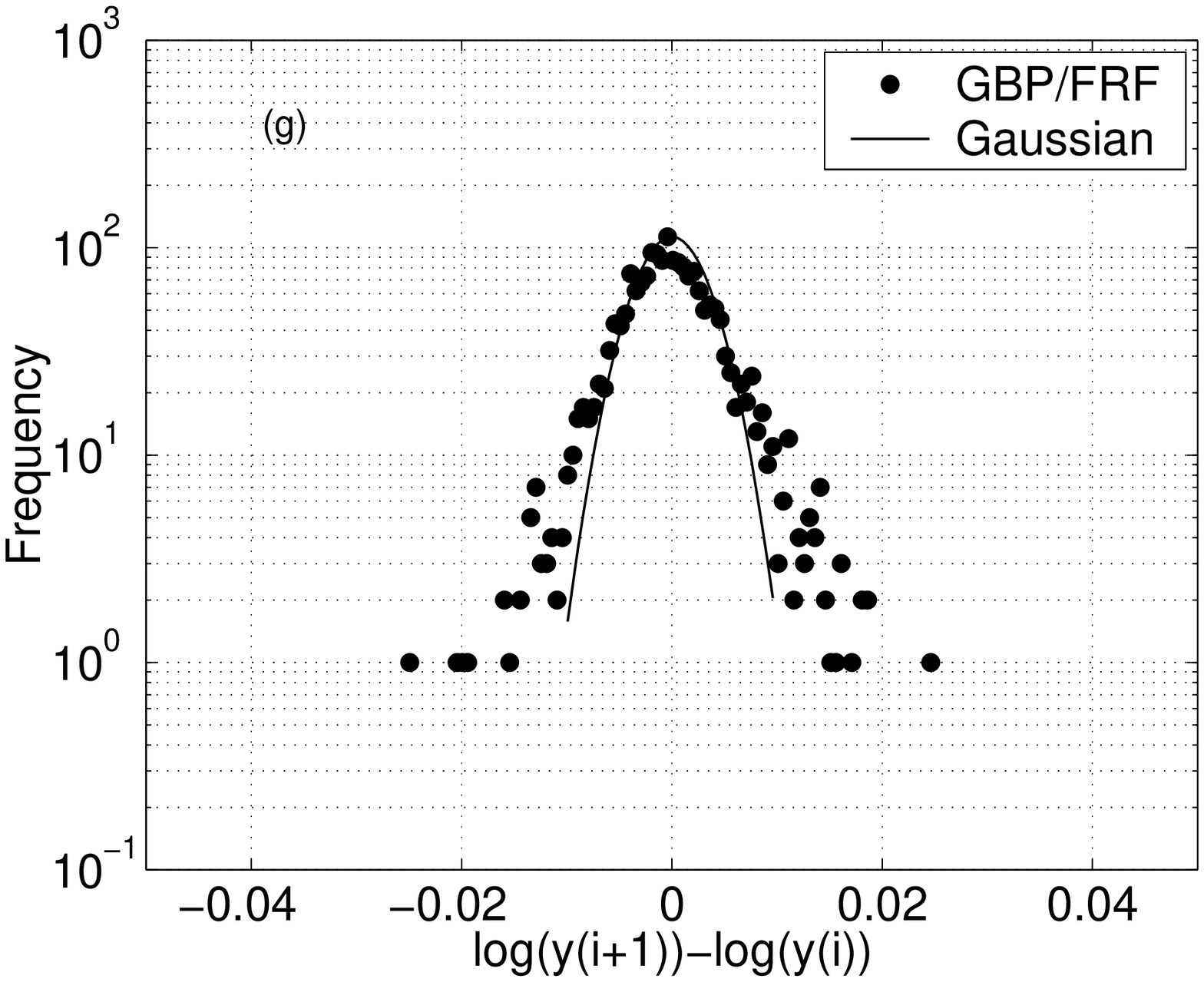} 
\hfill \includegraphics[width=.30\textwidth]{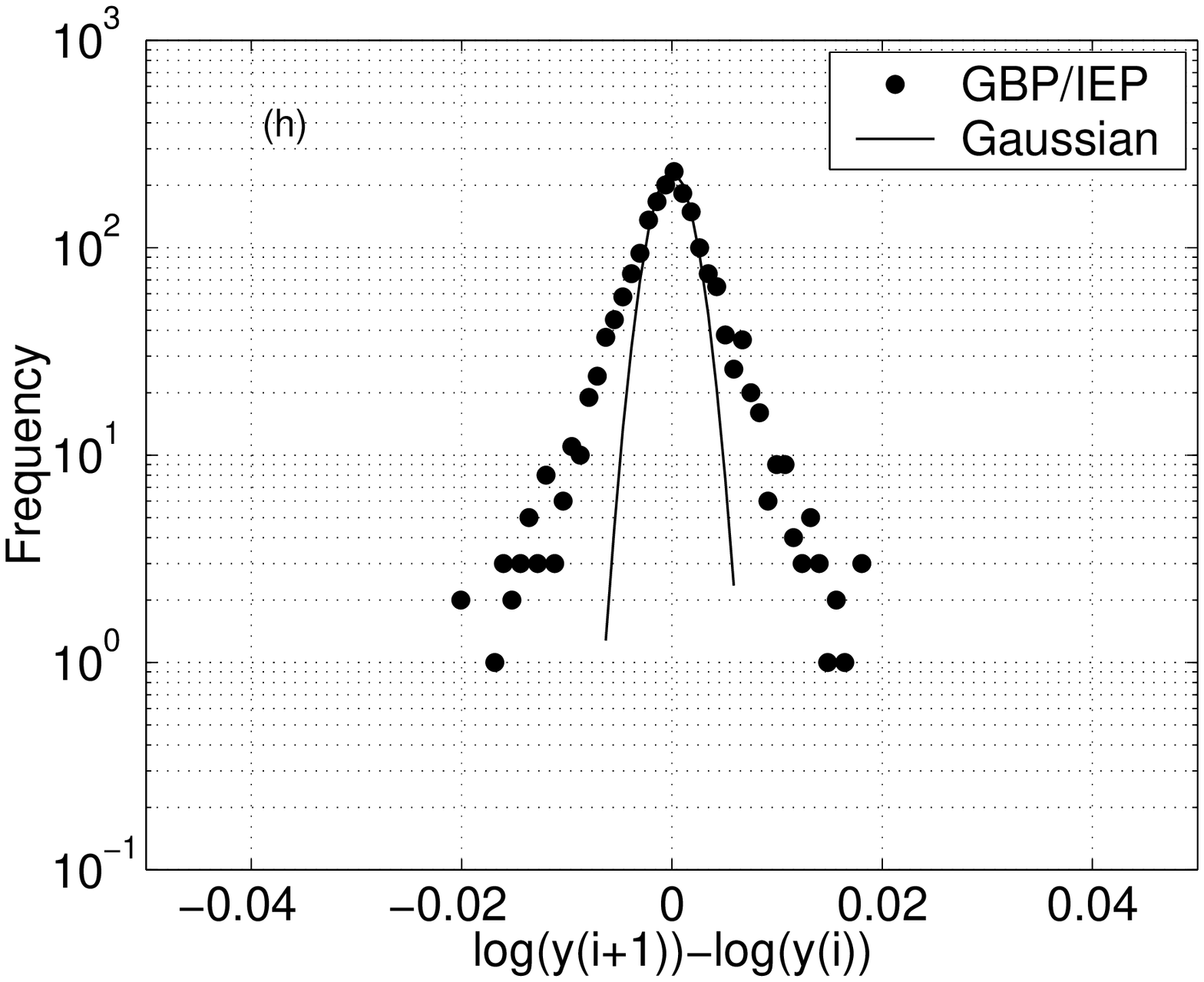} 
\vfill \includegraphics[width=.30\textwidth]{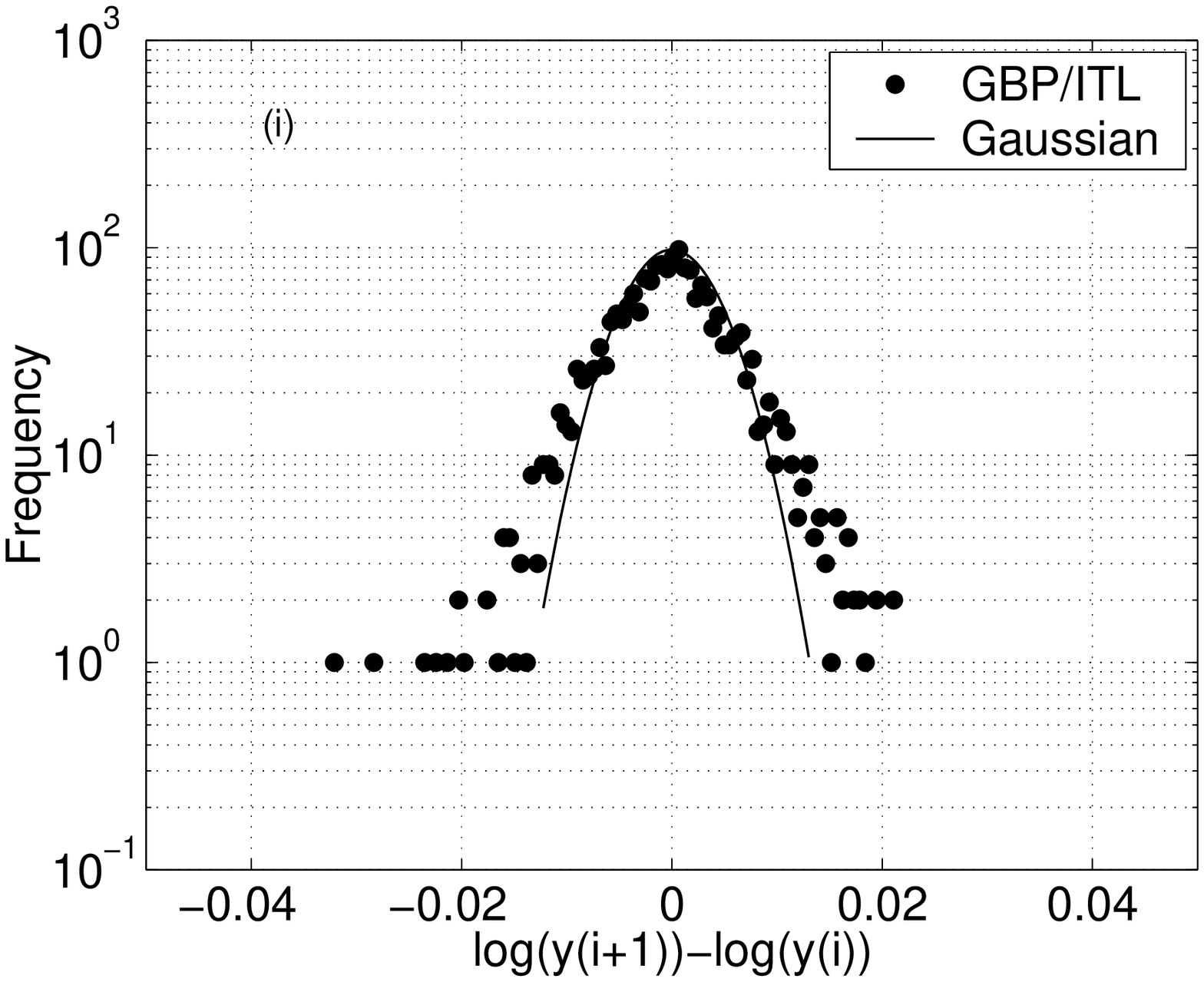} 
\hfill \includegraphics[width=.30\textwidth]{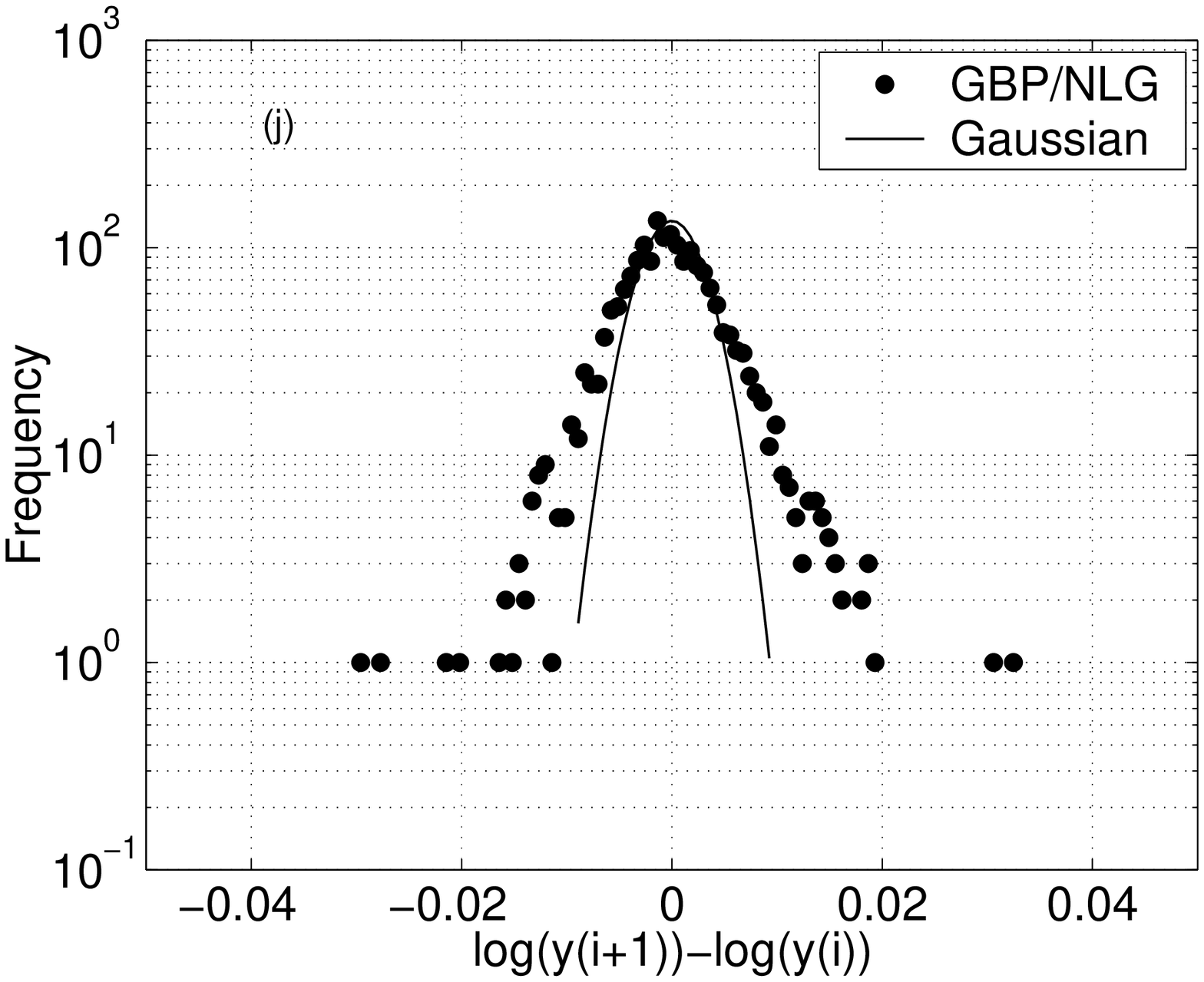} 
\hfill \includegraphics[width=.30\textwidth]{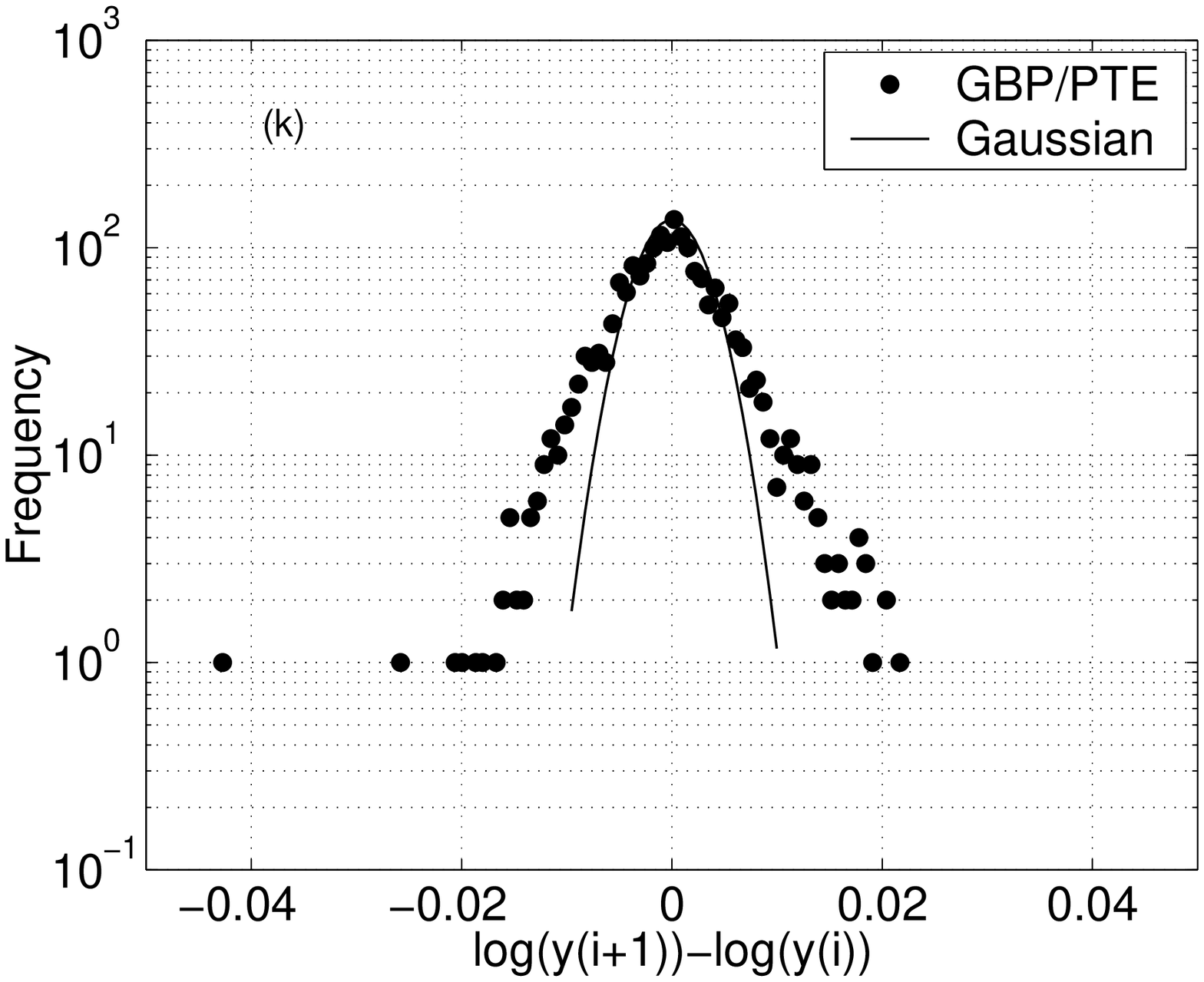} 
\caption{Distributions of the exchange rate
fluctuations for (a) $EUR/GBP$, and (b)-(k) $EUR/C_i$ for data in Fig. 1.}
\label{eps2} \end{figure}

\section{Correlations between fluctuations \\in $GBP$ exchange rates}

The $DFA$ technique\cite{DFA} leads to investigate whether the root-mean-square
deviations of the fluctuations of an investigated signal $y(n)$ have a scaling
behaviour, i.e. whether the $DFA$ function

\begin{equation}  <F^2(\tau)>   = \langle {1 \over \tau }
{\sum_{n=k\tau+1}^{(k+1)\tau} {\left[y(n)- z(n)\right]}^2} \rangle \sim
\tau^{2\alpha} \end{equation}

\noindent scales with time; $z(n)$ is a linear (trend) function fitting at best
the data in the $\tau$-wide interval which is considered. A value 
$\alpha = 0.5$
corresponds to a signal mimicking a Brownian motion.

A log-log display of the $DFA$ function leading to a measure of 
$\alpha$ for the
11 exchange rates of interest is found in the inset of Fig. 3. The
$\alpha$-exponent values for each national currency are summarized in Table 1.
The time scale invariance holds from 5 days (one week) to about 250 days ($ca.$
52 weeks or one banking year) showing a Brownian-like type of correlations.

\begin{figure} \centering \includegraphics[width=.8\textwidth]{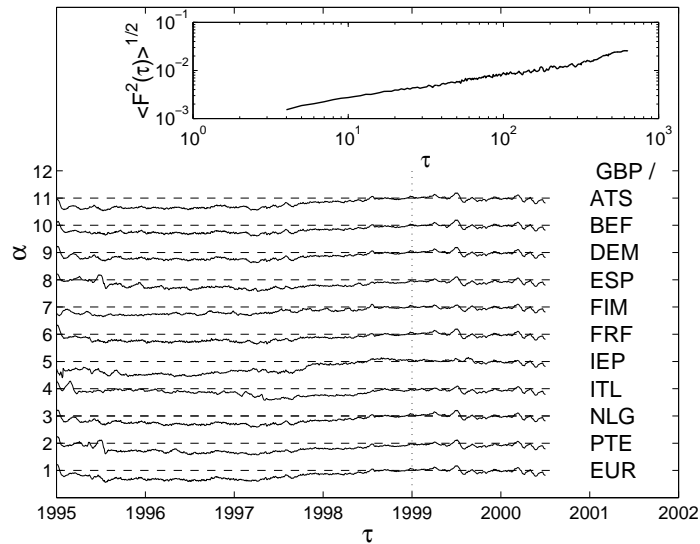}
\caption{Time dependence of the DFA $local$
$\alpha$-exponent for $EUR$ and each currency (which forms the $EUR$) exchange
rate with respect to $GBP$. The $\alpha$-values are artificially multiplied by
two and then displaced along the vertical axis in order to make the 
fluctuations
noticeable.  Insert : Log-log plot of the DFA function showing how to 
obtain the
$\alpha$ exponent for the 11 exchange rates of interest for $EUR/GBP$.}
\label{eps3} \end{figure}

In all cases, notice that the value of $\alpha$ for the $EUR$ ExR falls very
close to the Brownian motion value, - a result indicating in 
particular the good
sense of creating and using the $EUR$, thus guarding against speculations like
those having existed on national european currencies.

In order to probe the existence of {\it locally correlated} and {\it
decorrelated} sequences, an observation box, i.e. a 514 days (two years) wide
window probe is placed at the beginning of the data, calculated 
$\alpha$ for the
data in that box, moved this box by one day toward the right along the signal
sequence, calculated $\alpha$ in that box, a.s.o. up to the $N$-th day of the
available data. A local, time dependent $\alpha$ exponent is thus found.

The time dependent $\alpha$-exponent for $EUR$/$GBP$ ExR are shown in Fig. 3,
where $\alpha$ is determined from the best fit over the central, from 11 to 67
days (i.e., from 2 to 13 weeks) box/interval maintained to be constant for any
day on the examined time evolution.

The evolution of fluctuation correlations in the currency ExR are 
very well seen.
Notice how different the $EUR$/$GBP$ ExR behaves with respect to $ITL$/$GBP$.
Interestingly the $\alpha$-exponents for $ESP$, $DEM$ and $FRF$ are 
very close to
the $EUR$-$\alpha$-exponent behaviour over the whole period, indicating the
''control'' of such currencies over the nine others. Notice the 
special behaviour
of a country currency quite tied to the UK, i.e. the behaviour of
$IEP$-$\alpha$-exponent is markedly different from the other $EUR$ partners.

In Fig. 4 the time evolution of the statistical mean, median and standard
deviation of the $\alpha$ exponents for the currencies forming the $EUR$, are
compared to that of the $EUR$ in the ExR to $GBP$. Since the media is 
sometimes a
better representation of the main behaviour of a system, it is of interest to
consider the ratio between $\alpha_{mean}$ and $\alpha_{median}$, 
i.e. the upper
curve in Fig. 4. The mean/median ratio has large fluctuations before 1996, but
clearly tends to 1 thereafter; except for some 1999 spring period, the ratio is
now a constant.

\begin{figure} \centering \includegraphics[width=.8\textwidth]{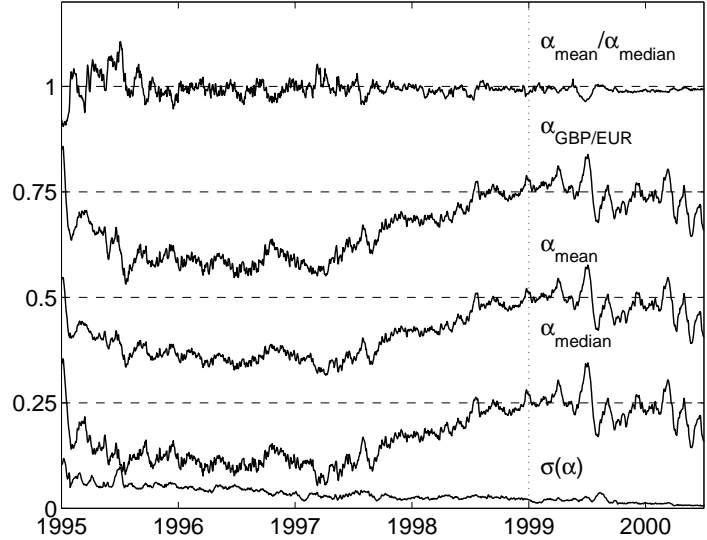}
\caption{Time evolution of the mean, 
median and
standard deviation of the $\alpha$ exponents for the currencies forming the
$EUR$, compared to that of $EUR$ ExR with respect to $GBP$ and the
$\alpha_{mean}/\alpha_{median}$ ratio. The $\alpha_{mean}$ and $\sigma(\alpha)$
curves are not displaced. The $\alpha_{median}$ curve is displaced by 
-0.25. The
$\alpha_{EUR/GBP}$ curve is displaced by +0.25. Horizontal dashed lines mark
Brownian motion 0.5 level for each $\alpha$-data.
}
\label{eps4} \end{figure}

\section{Intercorrelations between fluctuations}

A graphical correlation matrix of the time-dependent $\alpha$ exponent has been
constructed for the various exchange rates of interest. In Fig. 5,
$\alpha_{C_i/GBP}$ {\it vs.} $\alpha_{EUR/GBP}$ are shown for all $i$ values.
This so-called correlation matrix is displayed for the time interval hereby
considered, i.e. from Jan. 01, 1995 till Dec. 31, 1998.\footnote{The 
interval is
so chosen because the latter has to be reduced on one hand at the lower end due
to the size of the testing window box, and at the upper end by the fact that
after Jan. 01, 1999 the 11 currencies are not independent any more since their
conversion rates are fixed within the $EUR$.}

\begin{figure} \centering \includegraphics[width=.36\textwidth]{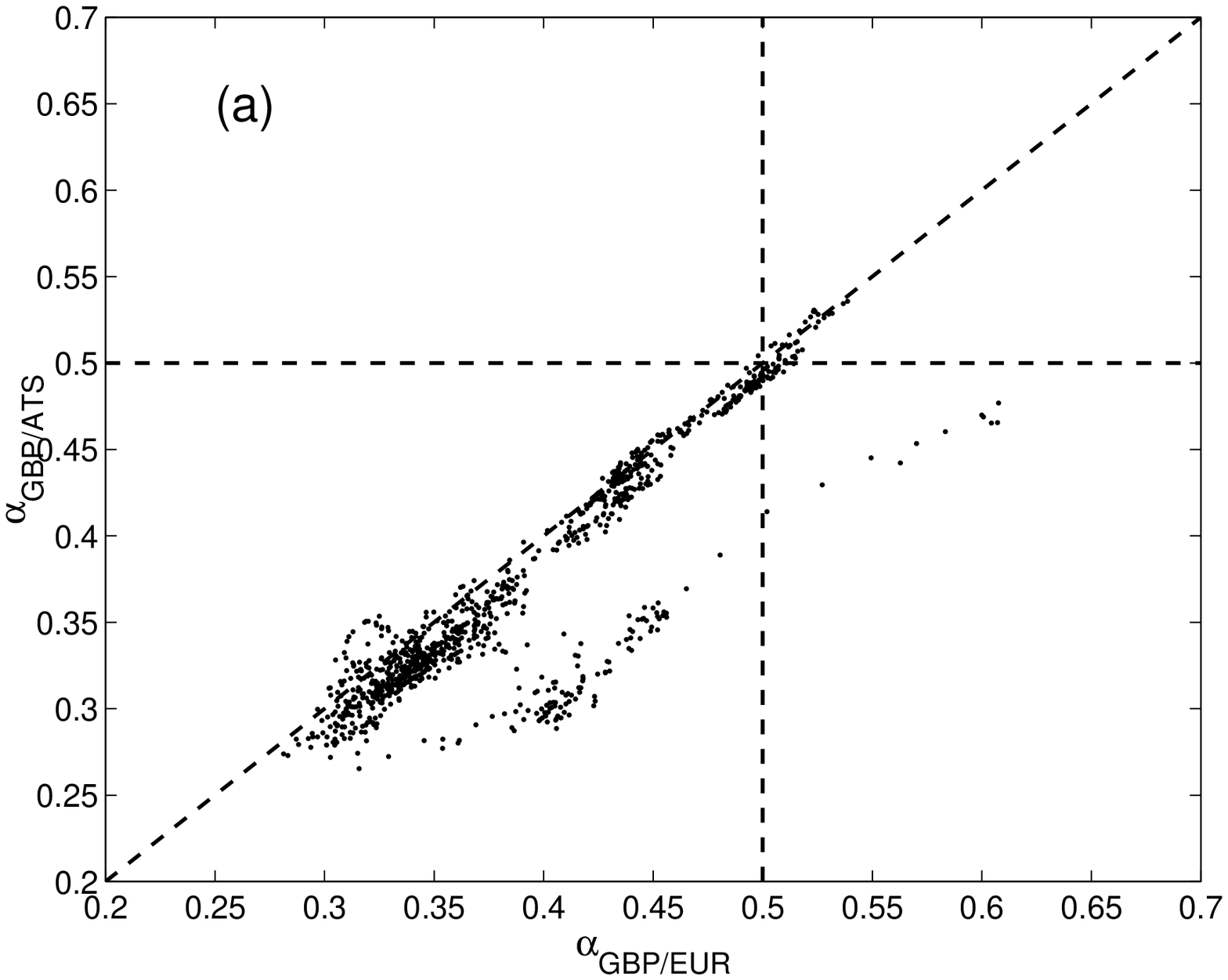}
\hfill \includegraphics[width=.36\textwidth]{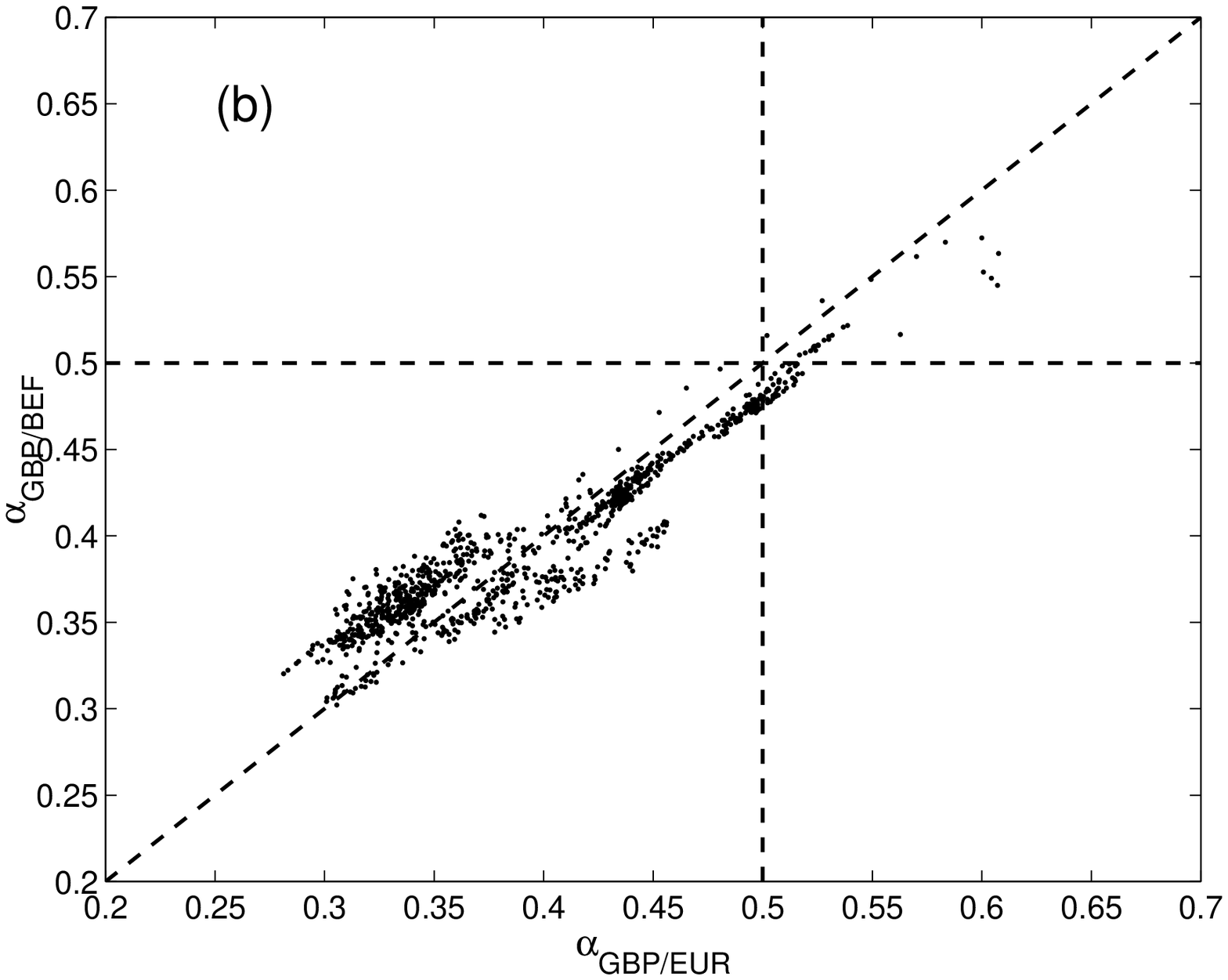} \vfill
\includegraphics[width=.36\textwidth]{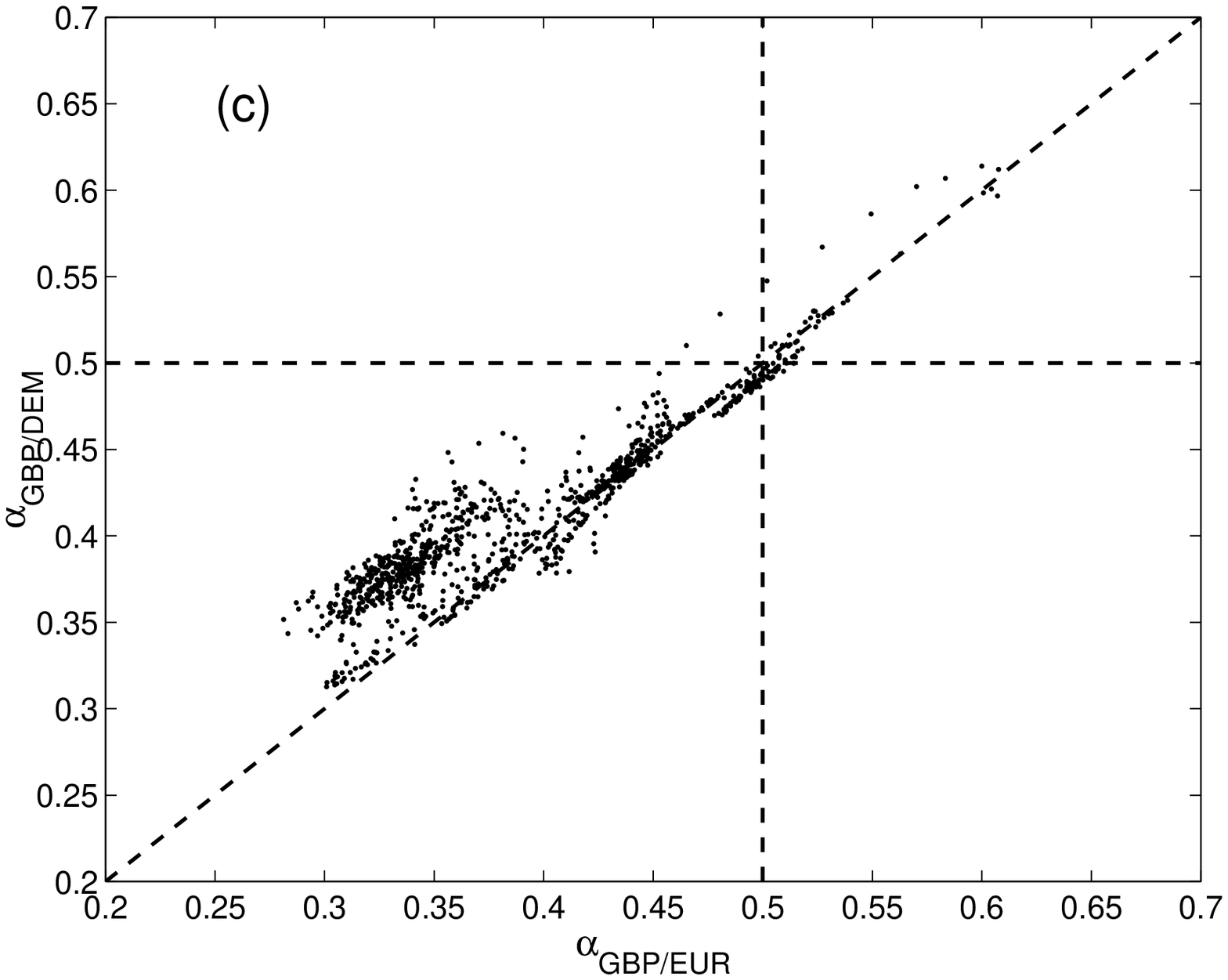} \hfill
\includegraphics[width=.36\textwidth]{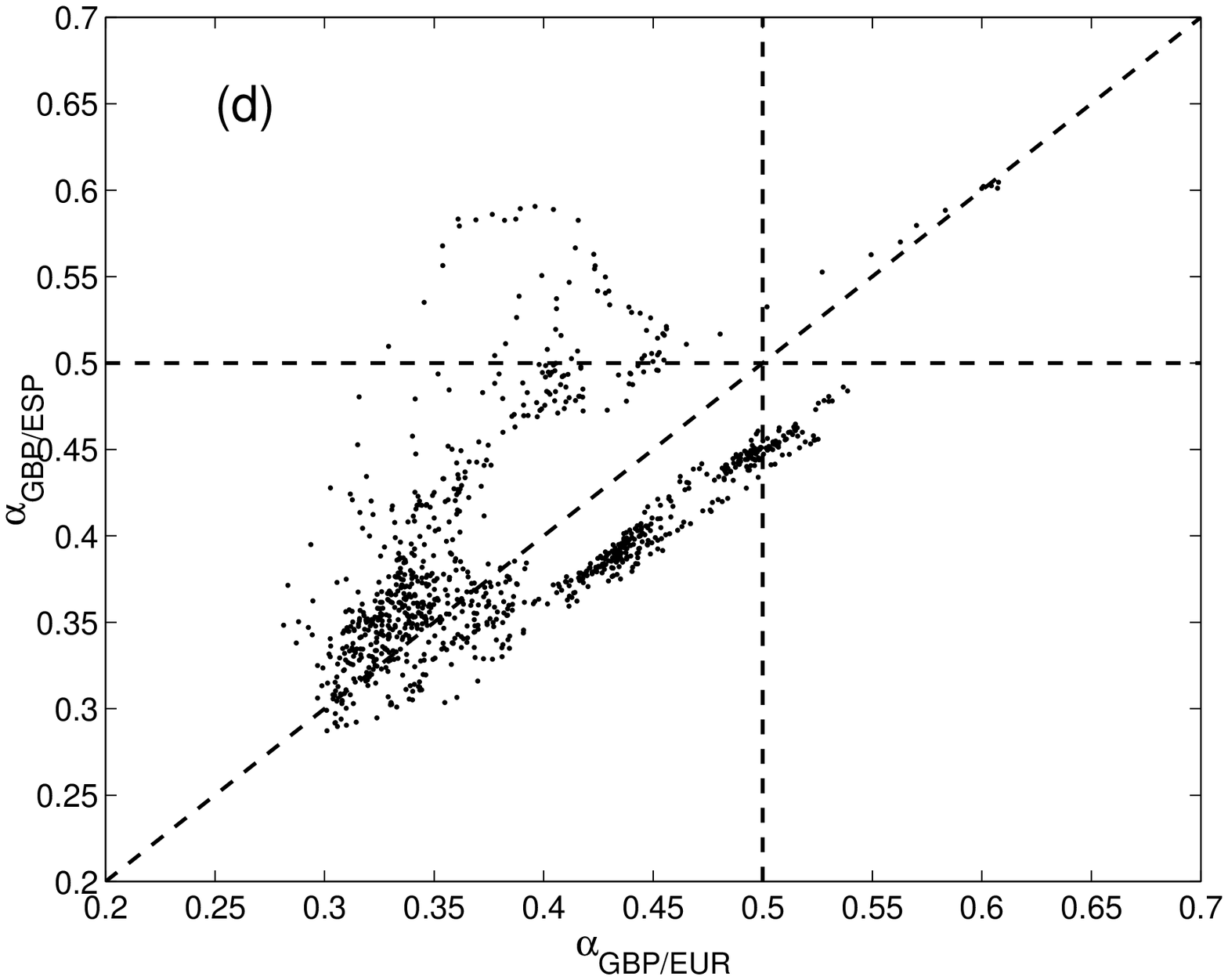} \vfill
\includegraphics[width=.36\textwidth]{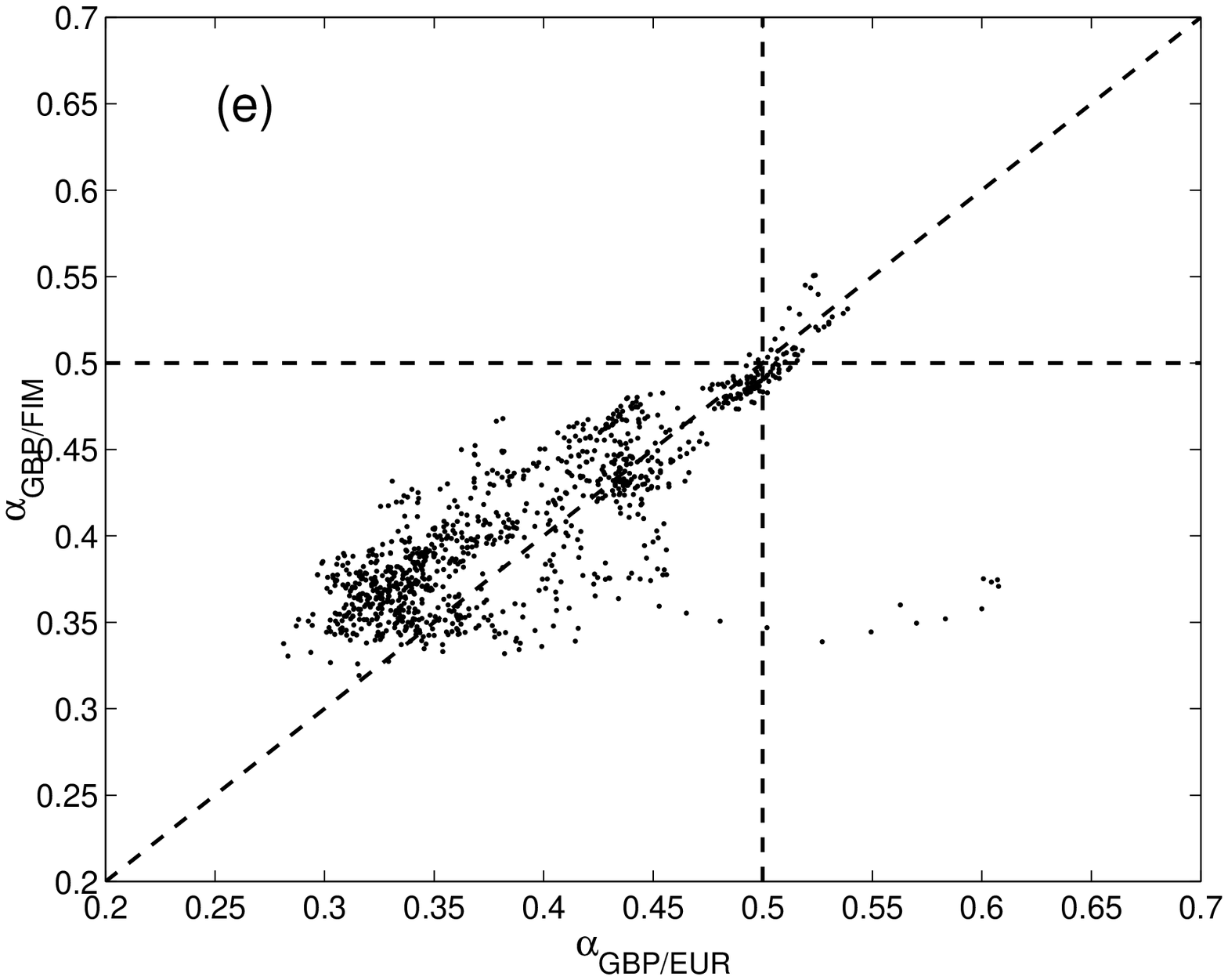} \hfill
\includegraphics[width=.36\textwidth]{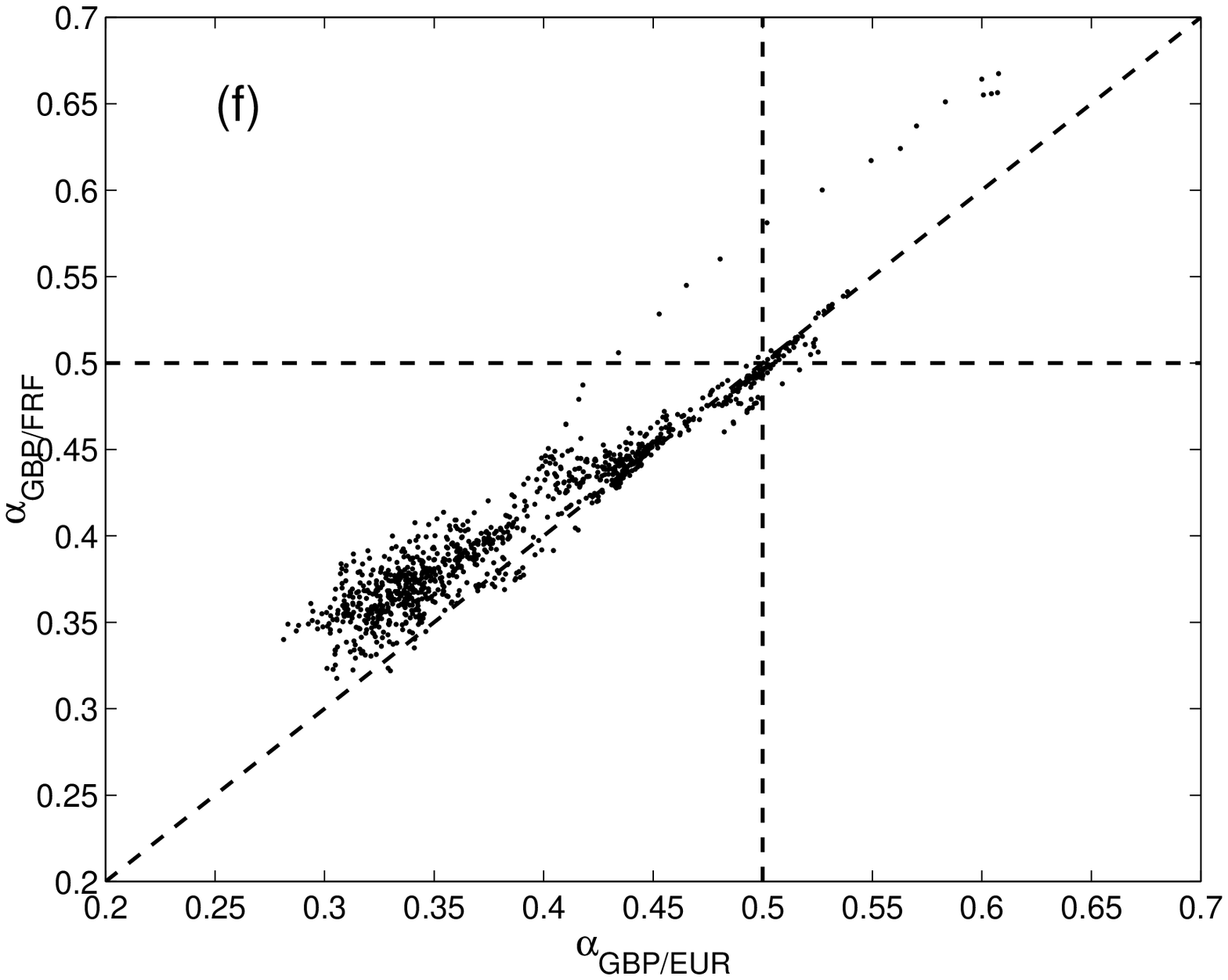} \vfill
\includegraphics[width=.36\textwidth]{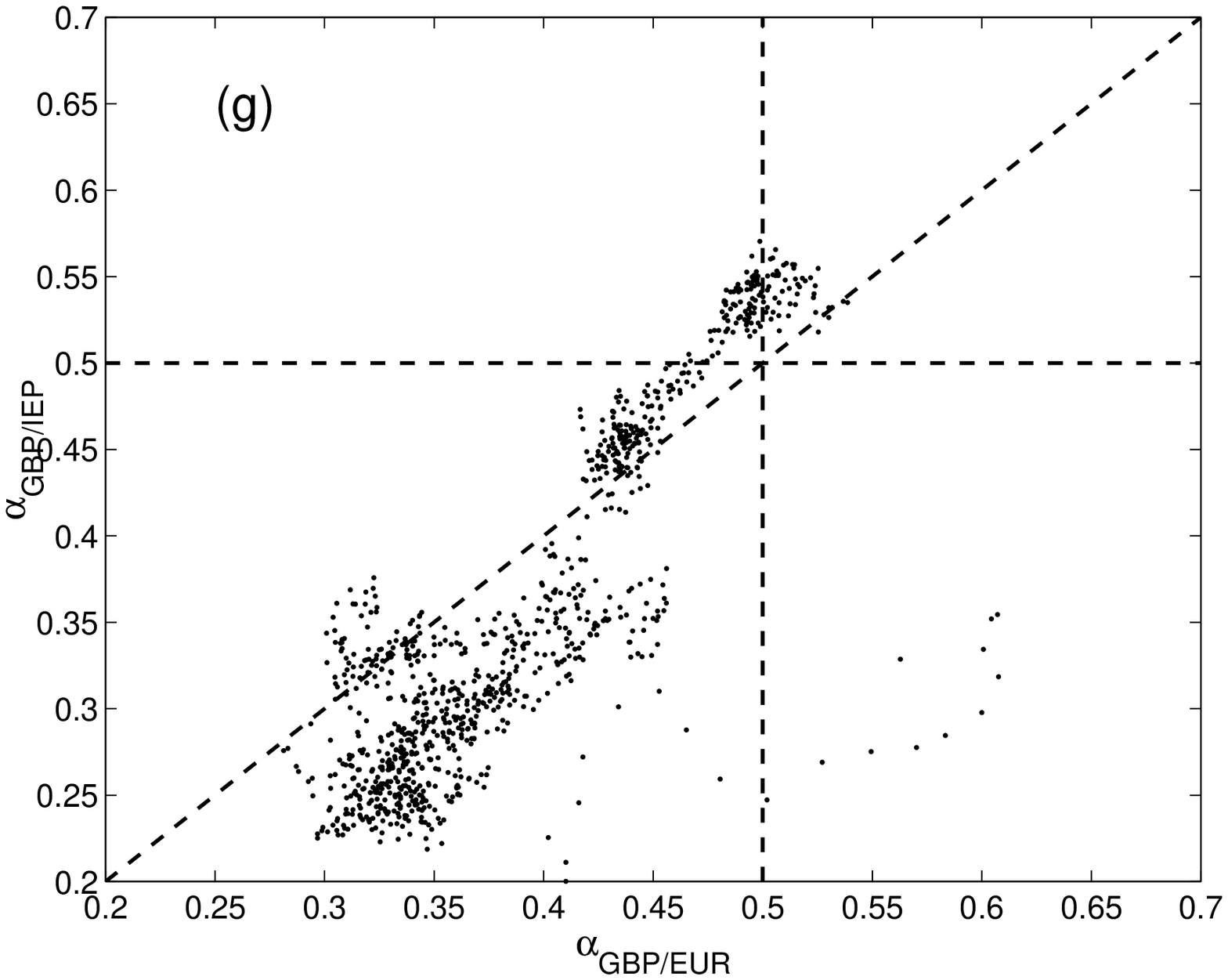} \hfill
\includegraphics[width=.36\textwidth]{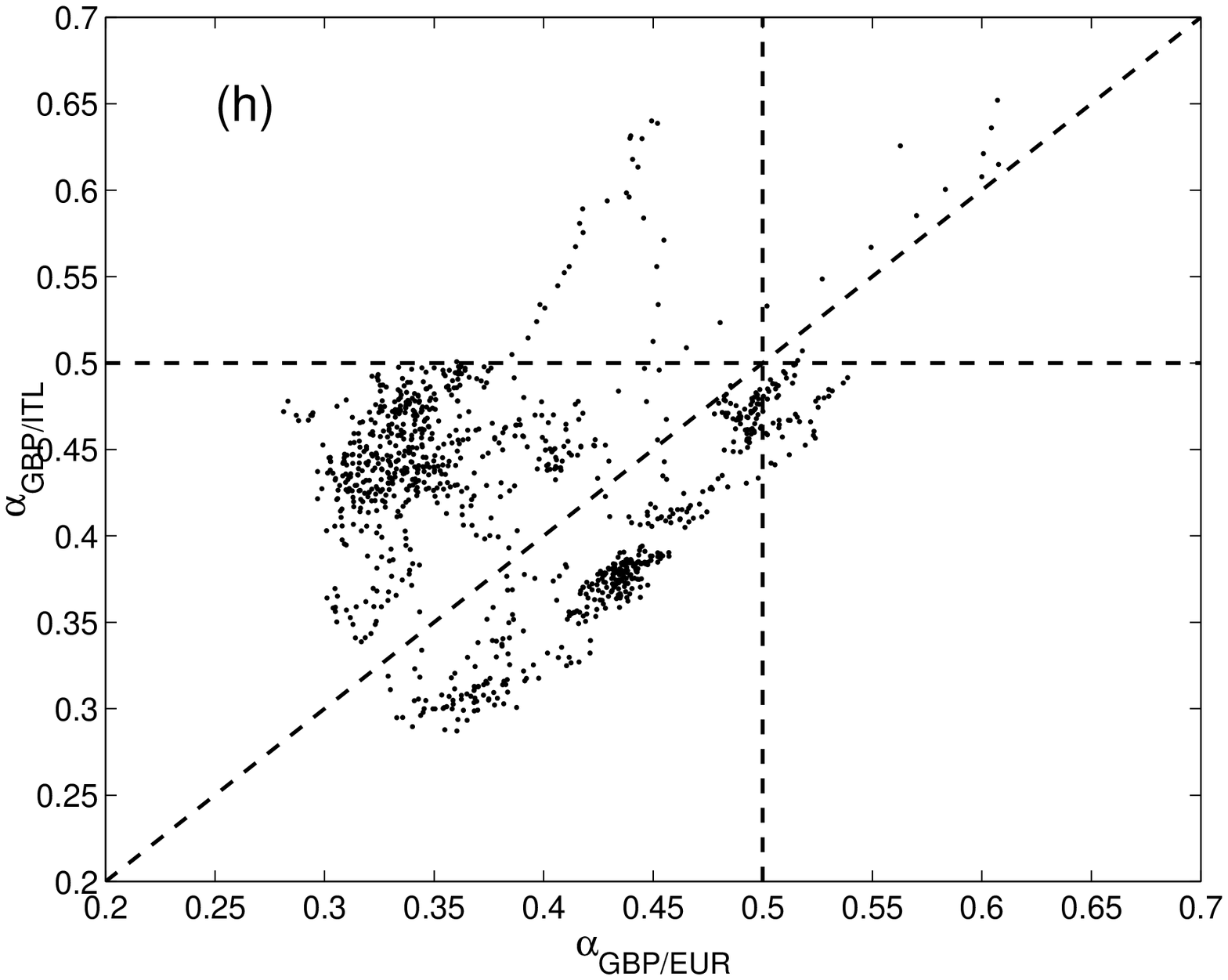} \vfill
\includegraphics[width=.36\textwidth]{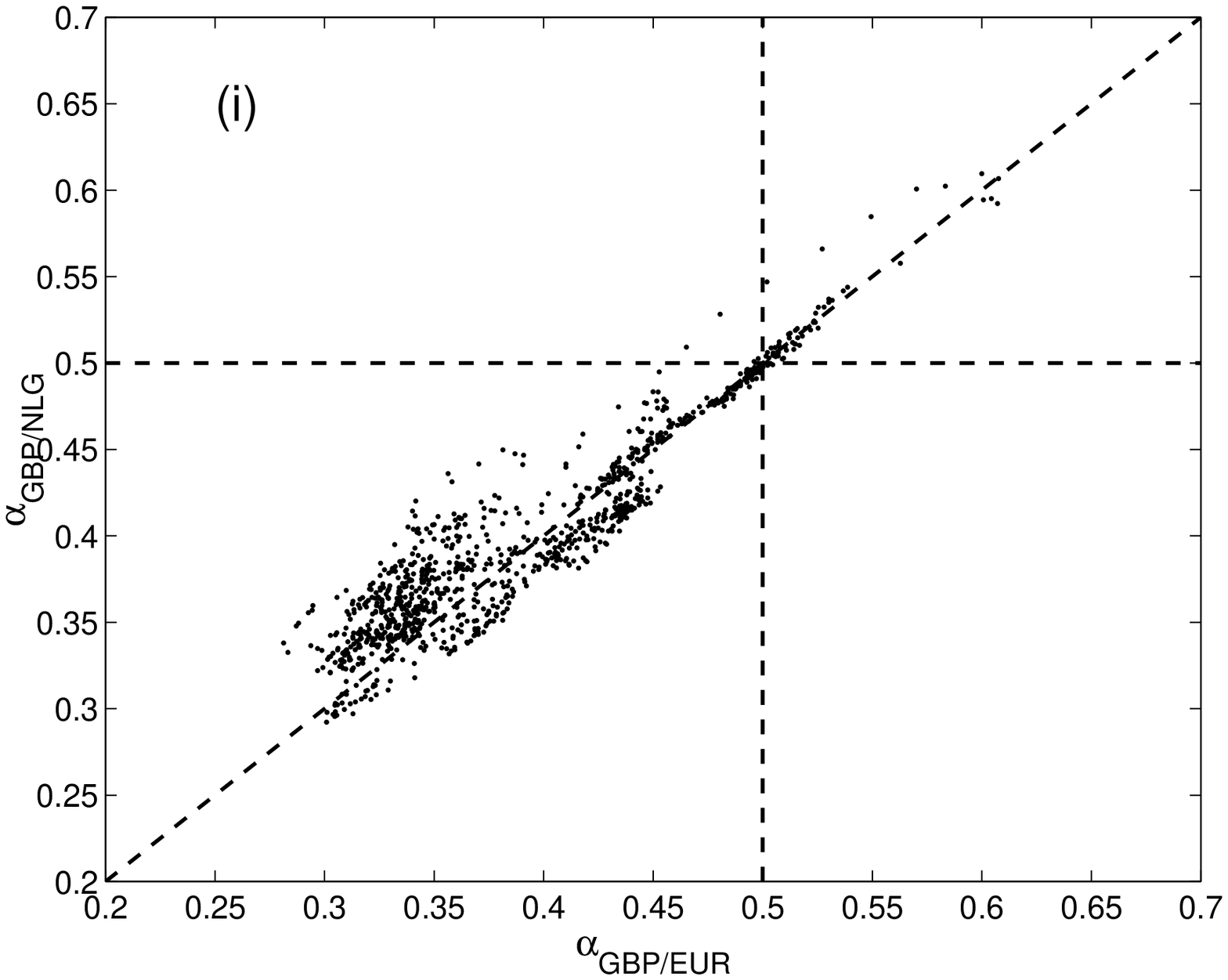} \hfill
\includegraphics[width=.36\textwidth]{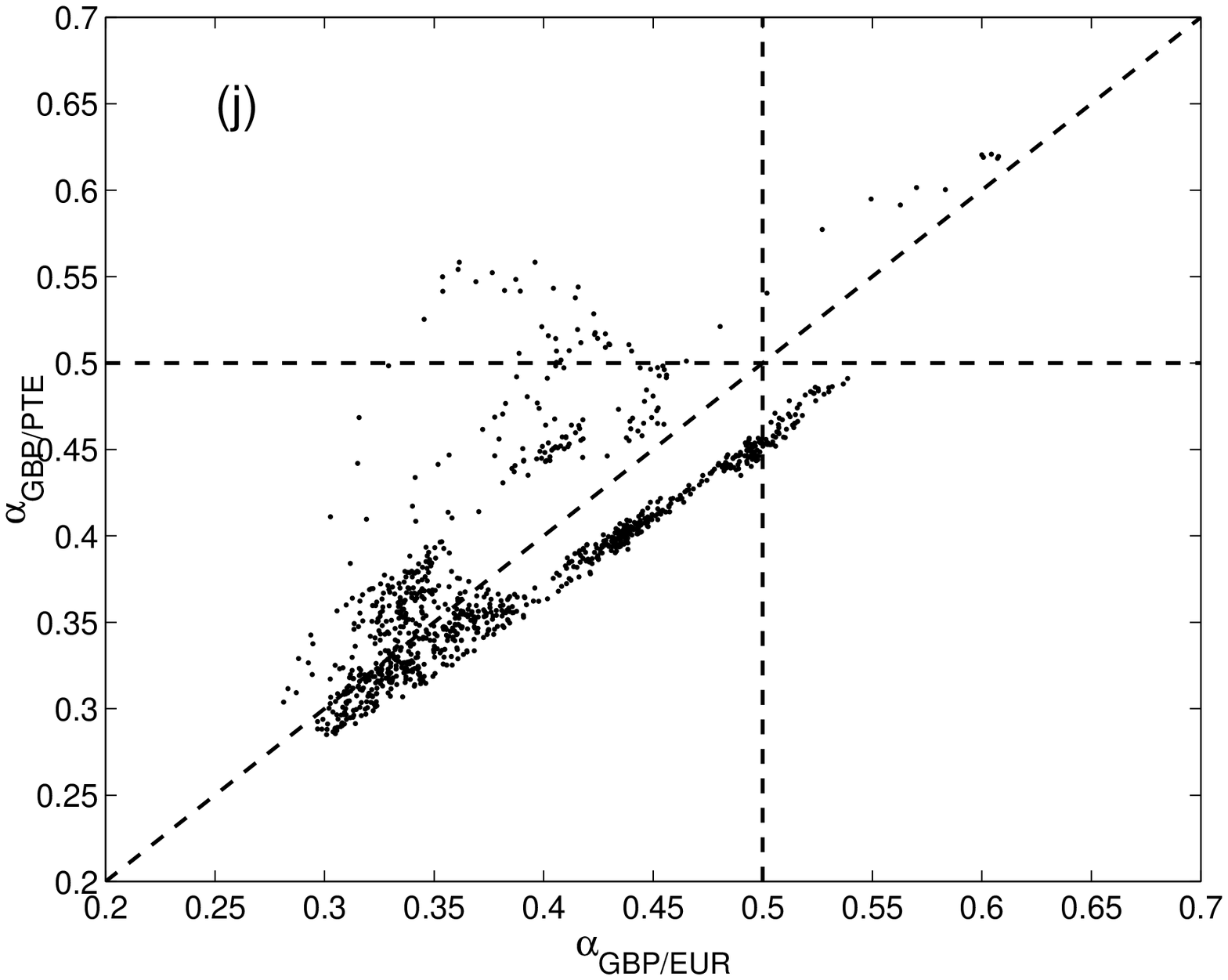} \caption{Graphical representation of the
so-called correlation matrix elements for the time interval Jan. 01, 1995 till
Dec. 31, 1999 for the various $local$ $\alpha_{C_i/GBP}$ {\it vs.}
$\alpha_{EUR/GBP}$ exponents, where $C_i$ are the ten $EUR$ currencies of
interest ($i$=1,10).} \label{eps5} \end{figure}

If the correlation is strong the cloud of points should fall along 
the slope $= +
1$ line. It appears that the about equally strong correlations exist 
between the
$EUR$/$GBP$ and the $DEM$/$GBP$, $EUR$/$GBP$ and $FRF$/$GBP$, $EUR$/$GBP$ and
$BEF$/$GBP$, $EUR$/$GBP$ and  $NLG$/$GBP$. Such a behaviour is related to the
fact that  $DEM$ and $FRF$ are the currencies of the leading economic 
countries,
while $BEF$ and $NLG$ are tightly related to both. Note that the structural
diagrams for $ITL$/$GBP$, $IEP$/$GBP$, and $ESP$/$GBP$ with respect to
$EUR$/$GBP$ show weak or no correlation at all.

\section{Conclusion}

A few aspects of the $EUR$ and its constitutive currencies exchange rates have
been studied from the point of view of the fluctuations of the exchange rates
toward $GBP$.

In examining various $reconstructed$ exchange rates for the currencies forming
the $EUR$ before Jan. 01, 1999 it has been searched whether correlations would
confirm historical and financial view points and so called standard knowledge.
The fluctuation distribution density as examined confirms that the foreign
exchange markets do not follow Gaussian distributions. The distribution of the
fluctuations is close to a Gaussian one only for small fluctuations, with
power-law distribution for large fluctuations. The correlation between
fluctuations were close to Brownian, and the more so after the $EUR$ was
introduced. There is no doubt that speculators have not found and do not find
some way to gamble on the $EUR$/$GBP$ exchange rates, since the DFA-$\alpha$
exponent is close to 0.5. It has been noticed that the introduction 
of the $EUR$
tends to smoothen the fluctuations and their correlations, as well gather
together in a main stream most of the European currencies, forming the $EUR$.

It is clear that the leading currencies from the point of view of the exchange
rate fluctuations are $DEM$ and $FRF$, with $ITL$ far away from the 
main stream.
The mean and median  $\alpha$ evolution for $EUR$/$GBP$ follows 
closely the mean
and median $\alpha$ exponents for the currencies in the $EUR$. This is in favor
of the conjecture that the $GBP$ has already been part of the $EUR$ system even
before Jan 1, 1999.

Thus will the $EUR$ be a gain for Britons or not? The answer is : not more than
now, on a statistical sense, since (i) $GBP$ is already a $EUR$ currency
according to the behaviour of local $\alpha$; (ii) the
$\alpha_{mean}$/$\alpha_{median}$ ratio is close to unity; (3) 
correlation matrix
elements are close to unity as well.

What will be the future of $GBP$? It will surely depend on the future of $EUR$,
and explanations, not counting hard statistical facts. According to a {\it
Referendum Street} program, broadcasted on Sunday Feb. 18, 2001 residents of a
North London borough, opposed euro entry by 65\% before hearing arguments; over
the week-end, hearing both side arguments, they were in favour by
58\%.\cite{crooks} Pro-euro had appealed to economic arguments, while the
anti-euro side stressed the loss of sovereignty and the existence of a European
super state. Our study shows that $GBP$ is already part of $EUR$.

\vskip 0.5cm

{\noindent \large Acknowledgements} \vskip 0.6cm This paper arises from a
question raised by P. Richmond (Trinity College, Dublin, and Norwich Union) at
Appllicatins of Physics to Financial Analysis (APFA2), Liège, Belgium, July
2000.\cite{kimalg}

\clearpage \addcontentsline{toc}{section}{Index} \flushbottom \printindex
%%%%%%%%%%%%%%%%%%%%%%%%%%%%%%%%%%%%%%%%%%%%%%%%%%%%%%%%%%%%%%%%%%%%%

\end{document}